\documentclass[12pt,titlepage]{article}
\usepackage{geometry}                % See geometry.pdf to learn the layout options. There are lots.
\usepackage{amssymb, amsfonts, amsmath}
\usepackage{graphicx}
\usepackage{helvet}
\usepackage[numbers,sort&compress]{natbib}

\newcommand{\e}{\mathrm{e}}

\newcommand{\tfa}{\ensuremath{{A}}}			% species A
\newcommand{\tfb}{\ensuremath{{B}}}			% species B
\newcommand{\Ta}{\ensuremath{{a}}}			%target A
\newcommand{\Tb}{\ensuremath{{b}}}			%target B

\newcommand{\LG}{\ensuremath{L_{\mathrm G}}}			%length of Genome
\newcommand{\Ec}{\ensuremath{E_{\mathrm{free}}}}		%E_cell
\newcommand{\Ens}{\ensuremath{E_{\mathrm{ns}}}}		%minim energy reduction when a TF binds DNA
\newcommand{\Eint}{\ensuremath{E_{\mathrm{int}}}}		%TF-TF interaction energy
\newcommand{\kT}{\ensuremath{k_{B}T\,}}		% kT unit
\newcommand{\Et}{\ensuremath{E_T\,}}			%Target energy
\newcommand{\Vc}{\ensuremath{V_{\mathrm{cell}}}}		%Volume Cell
\newcommand{\Vtf}{\ensuremath{V_{\mathrm{TF}}}}			%Volume TF
\newcommand{\Ltf}{\ensuremath{L}}	                % # of contacted basepairs by TF (=length TF)
\newcommand{\w}{\ensuremath{\omega}}			%cooperativity
\newcommand{\Na}{\ensuremath{N_{\tfa}}}
\newcommand{\Nb}{\ensuremath{N_{\tfb}}}
\newcommand{\N}{\ensuremath{N}}
\renewcommand{\S}{\ensuremath{S}}

\newcommand{\Eia}[1]{\ensuremath{E_{#1}^{\tfa}\,}}	%specific binding energy 
\newcommand{\Eib}[1]{\ensuremath{E_{#1}^{\tfb}\,}}
\newcommand{\mua}{\ensuremath{\mu_{\tfa}\,}}		%chem  potential A
\newcommand{\mub}{\ensuremath{\mu_{\tfb}\,}}		%chem  potential B
\newcommand{\fold}{\ensuremath{\phi\,}}	    	%foldchange

\newcommand{\ka}{\ensuremath{k_a}}	                %bimolecular reaction rate
\newcommand{\ksl}{\ensuremath{k_{\mathrm{sl}}}}	        %intrinsic hopping rate on DNA
\newcommand{\Zt}{\ensuremath{Z_{tot}}}				%total part function
\newcommand{\Zback}{\ensuremath{Z_{back}}}			%background (DNA+Cell)
\newcommand{\Zd}{\ensuremath{Z_d}}				%total DNA partition funciton
\newcommand{\Zc}{\ensuremath{Z_c}}				%Solvent partition function
\newcommand{\qns}{\ensuremath{q_{\mathrm{ns}}}}		%stat. weight binding ns-site
\title{Physical limits on cooperative protein-DNA binding and the kinetics of combinatorial transcription regulation}
\author{Nico Geisel\\
		Departament de Fisica Fonamental, Facultat de Fisica,\\
		 Universitat de Barcelona, Barcelona, Spain \\
    \and Ulrich Gerland\\
    Arnold-Sommerfeld Center for Theoretical Physics\\
    and Center for Nanoscience (CeNS), \\
    Ludwig-Maximilians-Universit\"{a}t, M\"{u}nchen, Germany
    }
    
\begin{document}
% generate the title page from the info in the headers above
\maketitle
% 200 words max Abstract
\abstract{ 
Much of the complexity observed in gene regulation originates from cooperative protein-DNA binding. While studies of the target search of proteins for their specific binding sites on the DNA have revealed design principles for the quantitative characteristics of protein-DNA interactions, no such principles are known for the cooperative interactions between DNA-binding proteins. We consider a simple theoretical model for two interacting transcription factor (TF) species, searching for and binding to two adjacent target sites hidden in the genomic background. We study the kinetic competition of a dimer search pathway and a monomer search pathway, as well as the steady-state regulation function mediated by the two TFs over a broad range of TF-TF interaction strengths. Using a transcriptional AND-logic as exemplary functional context, we identify the functionally desirable regime for the interaction. We find that both weak and very strong TF-TF interactions are favorable, albeit with different characteristics. However, there is also an unfavorable regime of intermediate interactions where the genetic response is prohibitively slow. 
}
\clearpage

\section*{Introduction}

Cells respond to many biochemical signals by adjusting their gene expression levels, often in a combinatorial way where the transcription rate of a given gene is a nonlinear function of several inputs. The entire signal transduction cascade, beginning with the detection of the biochemical signals and culminating in a changed intracellular protein concentration, is generally believed to be under strong selective pressure for rapid and well-adjusted responses in competitive environments. An important step in this cascade involves proteins belonging to the large class of transcription factors (TFs) which convey the external signal and trigger the appropriate genetic response by binding to specific binding sites on the genomic DNA. The search process of individual TFs for their functional target sites hidden within millions of non-functional sites on the DNA is well characterized, see e.g. \cite{Berg_Biochemistry_81, Bruinsma_PhysicaA_02, Halford_NAR_04, Slutsky_BiophysJ_04, Coppey_BiophysJ_04, Lomholt_PRL_05, Hu_BiophysJ_06}. This has led to an understanding of the tradeoffs inherent in the choice of TF-DNA interaction parameters, when both a rapid search as well as sufficient equilibrium discrimination for the functional sites is required \cite{vonHippel_PNAS_1986, Stormo_TIBS_98, Gerland_PNAS_02}. 

However, the experimental timescale for the search process, as inferred e.g. from single-molecule measurements {\it in vivo} \cite{Elf_Science_07}, is surprisingly short compared to the timescale for significant change in gene expression levels: Whereas a TF target site is occupied within a minute even at low TF concentrations, the concentration of the protein expressed from the target gene typically changes significantly only over a timescale of several minutes, due to the slow kinetics of protein synthesis and degradation. Hence, the search time is only a fraction of the total response time, and it is unclear whether fine-tuning of TF-DNA interaction parameters is needed for kinetic reasons. On the other hand, even in bacteria many genes are co-regulated by a combination of different TFs \cite{Richet_Cell_1991, Gerlach_JBact1991, Jennings_MolMicroBio_1993, Scott_MolMicroBio_1995, Pedersen_MolMicroBio1995, Brikun_JBact1996, Richet_EMBO_2000, Wade_EMBO_2001, Shin_EMBO2001}, while the search process studied so far is that of a single TF species, i.e. multiple TF molecules of the same type. A salient question is whether the timescale of transcription control increases with the complexity of the implemented regulatory function. 

To explore this question, we consider a simple theoretical model for the kinetics of combinatorial transcription regulation. We focus on the example of an AND-like cis-regulatory function implemented by two TFs, referred to as `$\tfa$' and `$\tfb$', which bind cooperatively to two adjacent target sites to activate a gene. This scenario is exemplified by the \textit{melAB} promoter of {\it E. coli}, where CRP and MelR bind cooperatively to activate transcription \cite{Wade_EMBO_2001}. Our model is sufficiently generic that it can be applied to a variety of cooperative protein-DNA binding situations. However, the example of the ``AND-gate'' is particularly well suited to illustrate the basic effects and functional tradeoffs that become apparent when the interaction parameters are varied. Compared to the well-studied case of a single TF-species, the new aspect here is the mutual interaction between the TFs (cf. Fig. ~\ref{fig1}), which is quantified by the dimensionless cooperativity $\w = e^{-\Eint/\kT}$. This quantity is only a measure of the interaction strength between TFs, with $\Eint$ the effective free energy of the interaction and $\kT$ the energy scale of thermal fluctuations. It is not related to the Hill coefficient, which depends on the number of components involved in a cooperative complex. The strengths of direct protein-protein interactions vary over a broad range with dissociation constants between the femto and the centi-molar regime \cite{Kumar_NucleicAcidsRes2006}. Biochemically feasible $\w$ values can therefore span many orders of magnitude, from weak transient interaction with $1<\w<1000$ to strong dimerization with $\w\sim10^7$ or larger. Depending on this value, the kinetics of cooperative protein-DNA binding will either be dominated by a ``monomer pathway'' or a ``dimer pathway'' \cite{Kohler_PNAS1999, Kohler_Biochemistry2001}. How do the response time and the steady-state levels of a regulatory module depend on the cooperativity? And which regime of $\w$ values could be favorable in which functional context? 

Our model, illustrated in Fig.~\ref{EnergyKinetics}, generalizes the classic facilitated diffusion model \cite{Berg_Biochemistry_81} to two interacting protein species. It incorporates the basic kinetic moves, i.e. binding to a DNA site, sliding along the DNA, and unbinding from the DNA, for monomers as well as for dimers. In addition, dimers can form or break up either in solution or while bound to the DNA. We characterize the behavior of our model using a variety of analytical and numerical approaches to calculate equilibrium and kinetic observables over a parameter range chosen to permit the exploration of functional tradeoffs in a bacterial system such as {\it E. coli}. For instance, in bacterial transcription regulation, a faster response is generally expected to be advantageous, whereas the steady-state transcription levels of a cis-regulatory function must be adjusted to yield the optimal protein concentrations for the biological conditions represented by the input signals \cite{Koch_JMolEvol1983, Dekel_NATURE_2005, Lang_PNAS2008}. Therefore, when considering different choices of $\w$, we compare regulatory systems that lead to the same steady-state levels. The exploration of our model leads us to two favorable regimes of $\w$, corresponding to weak (and often promiscuous) interactions and very strong heterodimerization, respectively. On the other hand, our model predicts that the search kinetics will be prohibitively slow at intermediate $\w$ values, at least when the protein copy number is small as is usually the case for bacterial transcription factors. In the `Discussion' section, we consider biological implications of these theoretical findings and discuss possible experiments to characterize the cooperative search problem and the kinetics of combinatorial transcription regulation.

%%%%%%%%%%%%%%%%%%%%%%%%%%%%%%%%%%%%%%%%%%%%%%%%
\section*{Results}
%%%%%%%%%%%%%%%%%%%%%%%%%%%%%%%%%%%%%%%%%%%%%%%%

\subsection*{Cooperativity and regulatory function}

Cooperative protein-DNA binding is employed in diverse functional contexts. 
For some functions, many molecules of the same protein polymerize along DNA, e.g. RecA for homologous recombination \cite{Galletto:2006p11196} or single-strand-binding-protein during DNA replication \cite{Lohman_ARB_94}. In these cases, the role of the protein-protein interaction is to enhance the probability of obtaining continuous DNA coverage rather than a patchwork of randomly positioned molecules. Here we focus on the functional context of transcription regulation where cooperative protein-DNA binding is involved in the processing of input signals. These signals are integrated and transformed into a single output, the transcription rate of a gene \cite{Ptashne_GenesSignals2001}. 

\begin{figure}[p]
\centering
\includegraphics[width=7.5cm]{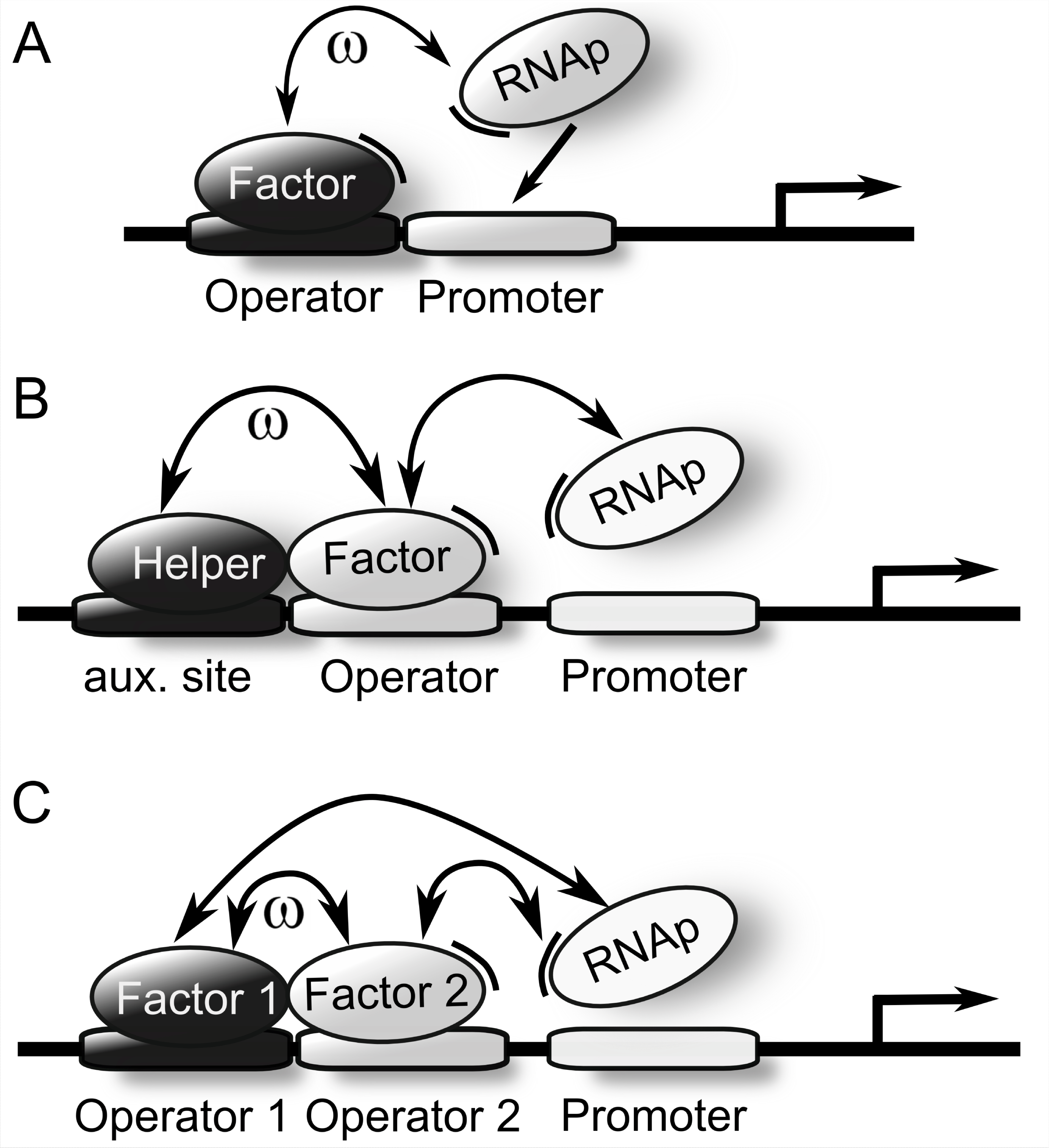}
\caption{Three schematic examples for cooperative protein-DNA binding in gene regulation.
In the signal transfer scenario (A) RNAp is recruited by an activating TF, whereby the signal conveyed by the TF is transferred to the transcription level. Scenarios (B) and (C) are both examples for signal integration. In scenario (B), an activator is assisted by a helper protein which does not contact RNAp itself. In scenario (C), two different TFs bind cooperatively and contact RNAp. These and other motifs are used by cells to implement regulatory functions \cite{Bintu_OPINION_2005a, Bintu_OPINION_2005b}, although the actual arrangement of TF binding sites in bacterial genomes is often more complicated, involving a larger number of sites \cite{Hermsen_PLOSCOMPBIO_2006}.}
\label{fig1}
\end{figure}

The cooperative binding of a transcription factor (TF) with RNA polymerase (RNAp) transfers a signal, by regulating the effective binding threshold for RNAp via the concentration of active TF (`regulated recruitment' \cite{Ptashne_GenesSignals2001}, see Fig.~\ref{fig1}A). When two different TFs bind cooperatively and each makes contact with RNAp to activate transcription, see Fig.~\ref{fig1}C, two signals are effectively integrated into a single output. A similar case is depicted in Fig.~\ref{fig1}B where TF binding is assisted by a helper protein that does not make contact with RNAp itself. This motif resembles, for instance, the regulation of the {\it melAB} promoter, which is co-dependent on the the transcription factors CRP and MelR \cite{Wade_EMBO_2001}. The helper can also be another molecule of the same TF, making the response to its signal more switch-like (increased effective Hill coefficient).  

The molecular function in the `signal transfer scenario' of Fig.~\ref{fig1}B is quantitatively described by the probability $p_{\Tb}$ to find a protein $\tfb$ bound as a function of the concentration of a protein $\tfa$ that binds adjacently. In contrast, for the `signal integration scenario', the functional activity is captured by the probability $p_{\Ta\Tb}$ that two DNA sites $\Ta$ and $\Tb$ are both occupied by the matching TF proteins. In the following, we will refer to both quantities, $p_{\Tb}$ and $p_{\Ta\Tb}$, simply as the `average activity' for the respective scenario. We envisage that selection acts on these average activities, as well as on a characteristic time scale, the `response time' $\tau$, associated with the kinetics of each mechanism. Here, $\tau$ corresponds to the typical delay for adjusting the activity to a new average level after a change in the input signal. In a steady state, $\tau$ is also a characteristic time scale of spontaneous fluctuations in the activity (noise). Importantly, both the average activity as well as the response time depend on the binding cooperativity $\w$.

\subsection*{Average activity}

Before we introduce our full model, it is instructive to consider the average activity within the simple approximation where we focus only on two binding sites $\Ta$ and $\Tb$ and ignore binding of the TFs to the rest of the DNA. This consideration will be useful in particular as a guide for our detailed study of possible tradeoffs in the choice of $\w$ within the full model. 

We first consider the signal transfer scenario as shown in Fig.~\ref{fig1}B. In equilibrium, the probability $p_{\Tb}$ that site $\Tb$ is occupied by one of $\Nb$ available molecules of type $\tfb$ is the normalized sum of the statistical weights for all states where $\Tb$ is occupied \cite{Bintu_OPINION_2005a}. In the absence of $\tfa$, i.e. for $\Na=0$, this is just $p_{\Tb}=q_{\Tb}/(1+q_{\Tb})$, with the statistical weight for an unoccupied site set to one and $q_{\Tb}=\Nb/n_{\Tb}$ denoting the relative weight for $\Tb$ to be occupied. Here, the `binding threshold' $n_{\Tb}$, which corresponds to the number of $\tfb$ molecules needed to obtain a $50\,\%$ average occupancy of $\Tb$ in the absence of $\tfa$, is directly connected to the effective equilibrium binding constant of $\tfb$ to $\Tb$ and the cell volume via $n_{\Tb}=K_{d}\Vc$. In the presence of $\tfa$, the occupancy of $\Tb$ increases to 
\begin{equation}
\label{eq-regulated-recruitment}
p'_{\Tb}=\frac{q'_{\Tb}}{1+q'_{\Tb}} \quad \mathrm{with} \quad q'_{\Tb}=q_{\Tb}\cdot \left[1+(\w-1)p_{\Ta}\right]\;,
\end{equation}
where $p_{\Ta}=q_{\Ta}/(1+q_{\Ta})$ is the average occupancy of $\Ta$ in the absence of $\tfb$. Thus, the presence of $\tfa$ boosts the statistical weight for $\tfb$ binding by the `regulation factor' \cite{Bintu_OPINION_2005a}, i.e. the factor in square brackets in Eq.~\ref{eq-regulated-recruitment}. Intuitively, this factor may be thought of either as a boost in the local effective concentration of $\tfb$ \cite{Ptashne_GenesSignals2001}, or as a decrease in the effective binding threshold $n_{\Tb}$ (the latter interpretation is closer to the underlying physics). 

Importantly, the regulation factor cannot exceed the cooperativity value $\w$, and it reaches $\w$ only if $p_{\Ta}$ takes on its maximal value of one. As a consequence, the cooperativity $\w$ is also an upper bound on the fold-change $\fold$ in $\Tb$-occupancy induced by a change in $\tfa$ concentration, since $p'_{\Tb}/p_{\Tb} \le q'_{\Tb}/q_{\Tb}$. This constitutes a physical constraint on $\w$ that arises from the equilibrium statistics of cooperative protein-DNA binding, 
\begin{equation}
\label{eq-omega-thermodynamic-constraint}
\w > \fold \qquad \mbox{[equilibrium constraint]},
\end{equation}
i.e. the cooperativity must be larger than the required fold-change $\fold$ in the output signal ($\fold=p'_{\Tb}/p_{\Tb}$ for the signal transfer scenario). On the molecular level, this constraint can be implemented by a sufficiently strong direct protein-protein interaction or by indirect mechanisms of cooperativity, e.g. via collaborative competition \cite{Miller_MolCellBiol2003} or DNA bending \cite{Spakowitz_PRL2009}. 

For the signal integration scenario in Fig.~\ref{fig1}C, the definition of the fold-change $\fold$ is different, but the constraint (\ref{eq-omega-thermodynamic-constraint}) holds as well. Here, the relevant fold-change is the average activity in the presence of both inputs relative to the average activity with only a single input, $\fold=p_{\Ta\Tb}/p_{\Ta}$ or $\fold=p_{\Ta\Tb}/p_{\Tb}$, where 
\begin{equation}
\label{eq-cooperative-activation}
p_{\Ta\Tb}=\w q_{\Ta}q_{\Tb}/(1+q_{\Ta}+q_{\Tb}+\w q_{\Ta}q_{\Tb})
\end{equation}
This fold-change is then transferred to the promoter activity in the example considered in Fig.~\ref{fig1}C. Taken together, when considering steady-state activities, both the signal transfer and the signal integration function benefit from larger cooperativities, since large $\w$'s allow for tight regulation. However, since large binding energies often lead to slow kinetics, we will explore whether a tradeoff exists between the fold-change in average activity and the response time.

\subsection*{Full model}

\begin{figure*}[p]
\centering
\includegraphics[width=8cm]{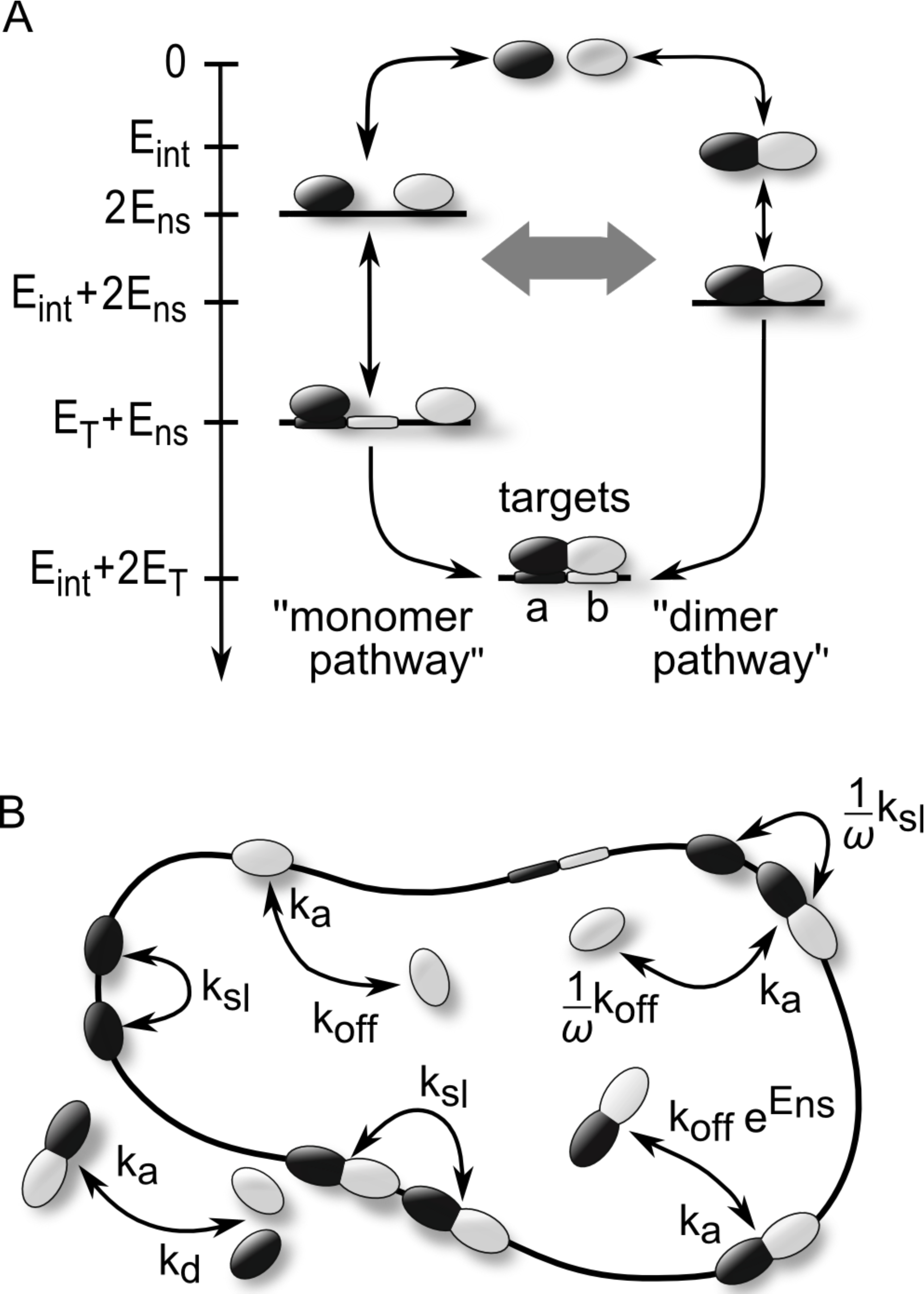}
\caption{Illustration of the energy levels and the kinetic model for the two TF species system with a non-specific genomic background. 
(A) Binding of TFs to the DNA reduces the energy by $\Ens<0$ compared to the unbound reference state with energy $\Ec=0$. Additional energy can be gained through sequence specific contacts (not shown). Upon dimerization of TFs in solution or on the DNA the energy is further reduced by the interaction energy $\Eint\leq0$. The TFs bind to their target site with a specific binding energy $\Et$. At small dimerization energies $\Eint$, full promoter activation will be reached via the ''monomer pathway'', where TFs arrive at their target independently and consecutively. At large $\Eint$, on the other hand, TFs will pre-dimerize in the DNA-background  or in solution and arrive to the targets simultaneously through the ''dimer pathway''. 
(B) Transcription factors dimerize in solution and bind to the DNA in diffusion limited binding reactions with a rate constant $\ka$. The dissociation rate of a free dimer $k_{\mathrm d}$ and the dissociation rate $k_\mathrm{off}$ of a TF from a DNA site depend on the corresponding energies and follow from detailed balance as explained in the main text. Dimers and monomers can randomly diffuse along the DNA with a rate $\ksl$, which becomes site dependent when the binding energy is sequence specific. When  the dissociation of a monomer requires the simultaneous dissociation from a cooperatively bound partner its off-rate $k_\mathrm{off}$ decreases by a factor $1/\omega$.
}
\label{EnergyKinetics}
\end{figure*}

We now introduce a full kinetic model for the cooperative target search which is based on the energies of TF binding states and the transition rates between these states, as illustrated in Fig.~\ref{EnergyKinetics}. We consider a single circular genome of length $\LG$ (in units of base pairs) inside a cell of volume $\Vc$ with a single pair of adjacent target sites for $\tfa$ and $\tfb$. The unbound state of free TFs in solution is our reference state, with its energy set to $\Ec=0$. If $\tfa$ and $\tfb$ dimerize in solution, the interaction energy $\Eint<0$ is gained, while entropy is lost, since the number of possible states is reduced by a factor that we write as $\Vtf/\Vc$, with a microscopic volume $\Vtf$ on the order of the size of a TF. Each TF molecule has $\LG$ possible binding sites on the DNA (indexed by $i$ with $0\leq i < \LG$) with the respective bound-state energies $\Eia{i}$ and $\Eib{i}$. These bound-state energies are either equal to $\Ens<\Ec$, if the protein-DNA interaction is non-specific, or they take on a lower value if the binding sequence favors specific protein-DNA contacts, $\Eia{i},\Eib{i}\leq\Ens$. We denote by $\Ltf$ the number of base pairs on the DNA which are occupied by a bound monomer (occupied DNA is inaccessible to other TF molecules), and we posit that $\tfa$ and $\tfb$ can form a DNA bound dimer only when $\tfb$ binds directly upstream of $\tfa$. 

For the kinetic rates, we assume that all binding reactions are diffusion limited. For simplicity, we take the same rate constant $\ka$ for the binding of two protein molecules in solution and for the association of a TF molecule with a specified DNA site (thus, the total rate of TF binding anywhere on the DNA is $\LG\ka$, if no DNA site is occupied already). The random diffusion of TFs along the DNA contour occurs with the basal sliding rate $\ksl$. When neighboring sites have different energies, the sliding rate is the basal rate $\ksl$ from the higher to the lower energy state while the reverse process occurs at the reduced rate $\ksl\exp(-\Delta E/k_{B}T)$, with $\Delta E>0$ the energy difference, such that detailed balance is respected (in the following we assume all energies to be in units of $k_{B}T$ which amounts to setting $k_{B}T=1$). The rates for all other possible reactions are similarly obtained from detailed balance. For instance, the unbinding rate $k_\mathrm{off}$ of a monomer from a non-specific DNA site is determined by $k_\mathrm{off}/\ka=(\Vc/\Vtf) e^{(-\Ec+\Ens)}$, and the dissociation rate $k_{\mathrm d}$ of a free dimer $k_{\mathrm d}/\ka=(\Vc/\Vtf) e^{\Eint}$. Note that monomers can also unbind or slide away from a DNA site while simultaneously dissociating from a cooperatively bound partner (thus disrupting the DNA-bound dimer, see Fig.~\ref{EnergyKinetics}b, top right). In that case detailed balance dictates that monomer sliding and dissociation rates decrease by a factor $1/\w$ due to the loss of the dimerization energy $\Eint$. 

Within the framework of this full model, we calculate the steady-state activities as described in Section S1 in the Supporting Material (this exact calculation includes the effect of the genomic background and mutual exclusion of overlapping binding sites, both neglected in the simple discussion above). We determine average search times numerically, using kinetic Monte Carlo simulations as described in Section S2, and we also develop analytical approximations further below and in Section S3. 

We choose the parameters of our full model to roughly reflect the situation in a bacterium such as {\it E. coli}. We set the genome length to $\LG = 5\cdot 10^{6}$~bp, choose a cell volume of $\Vc=5\,\mu{\mathrm m}^3$, and consider DNA binding sites of length $\Ltf=15$~bp. The sliding rate $\ksl$ can be determined from recent measurements of the one-dimensional diffusion constant for TF sliding on non-specific DNA \cite{Elf_Science_07, Wang_Biophys_2009}, which obtained values close to $0.05 \mu {\mathrm m}^ 2/{\mathrm s}$, corresponding to a sliding rate of about $\ksl = 10^5/{\mathrm s}$. The same experiments also determined a residence time of $0.3-5$~ms for TF molecules on non-specific DNA before dissociation. At the given genome length, this fixes our rate constant $\ka$ to be in the range $0.4-6\cdot 10^{-3}/{\mathrm s}$, and we set $\ka=10^{-3}/{\mathrm s}$ in the following. Unless otherwise stated, we will assume, for simplicity, that the target sites $\Ta$, $\Tb$ are the only specific binding sequences in the genome, both with energy $\Et$. We measure all energies in units of $k_BT$. We set the strength of the non-specific protein-DNA interaction by requiring that a single TF spends on average equal time unbound in solution as bound somewhere on the DNA. This parameter choice corresponds to the well-characterized optimum for the search process of a single TF species \cite{Winter_Biochemistry_81b, Slutsky_BiophysJ_04}; see also the discussion of this point further below. Within our energy model, this corresponds to a non-specific binding energy $\Ens=\log\left(\LG\cdot\Vtf/\Vc\right)=-5.3$, assuming a reaction volume $\Vtf=1$~nm$^3$. 
In our model, the effective dimerization rate is increased by the presence of the DNA (which acts as a scaffold for the interaction). A similar increase was observed experimentally in a study of the Jun-Fos DNA complex \cite{Kohler_Biochemistry2001}.

\subsection*{Quantitative Analysis}

We now analyze how the quantitative characteristics of the two-protein-species system depend on the cooperativity $\w$. The cooperative target state where both target sites are occupied can be reached via two distinct kinetic pathways: In the ``monomer pathway'', $\tfa$ and $\tfb$ separately search for their specific target sites in multiple rounds, alternating between one-dimensional diffusion along the DNA and three-dimensional diffusion in the cytoplasm to a new position on the DNA. In this pathway, $\tfa$ and $\tfb$ arrive independently, i.e., one after the other, at their specific target sites. By contrast, in the ``dimer pathway'', the dimer forms beforehand, either in solution or in the DNA background, such that $\tfa$ and $\tfb$ reach their target sites simultaneously (cf. Fig.~\ref{EnergyKinetics}A). Clearly, we expect the monomer pathway to dominate for weak TF-TF interactions (small $\w$), while the dimer pathway should dominate for large $\w$. But what is the behavior of the overall search time $\tau$ that results from the kinetic competition between the two pathways? 

Before performing the kinetic analysis, we first characterize the steady state characteristics of our full model. We will focus on the signal integration scenario in the remainder of this study; the behavior in the signal transfer scenario is qualitatively similar. As discussed above, the most relevant steady state characteristic in the functional context of gene regulation is the attainable fold-change of the average activity, which determines how tightly a gene can be regulated. We assume that the expression level of the regulated gene in the high-activity state, when both TF species can bind the promoter (the ``ON-state'') is constrained to its optimal level by evolutionary selection, e.g. the optimal level of a metabolic enzyme in the presence of its substrate \cite{Dekel_NATURE_2005, Koch_JMolEvol1983}. The fold-change between the ON-state and the OFF-state (in which only one of the TFs can bind) then determines how tightly the production of the protein can be suppressed under conditions when it would be useless or even detrimental. Hence, when we consider the system at different cooperativity values $\w$, we take for granted that another system parameter is adjusted to keep the ON-state level constant. Specifically, we will assume that this compensation occurs via the target binding threshold, which is programmable via the DNA sequence of the target site \cite{Gerland_PNAS_02}. In other words, we compensate a weaker protein-protein interaction with a stronger protein-target interaction such that the ON-state level $p_{\Ta\Tb}$ remains constant. In E.coli and yeast, binding sites indeed tend to deviate from the consensus motif when multiple TFs bind next to each other in the cis-regulatory region \cite{Bilu_GenomeBiol2005, Richet_EMBO_2000, Scott_MolMicroBio_1995}. For simplicity, we consider a symmetric pair of TFs, which have different binding sequences, but the same energetics, such that $q_{\Ta}=q_{\Tb}$.

\begin{figure}[p]
\centering
\includegraphics[width=8.6cm]{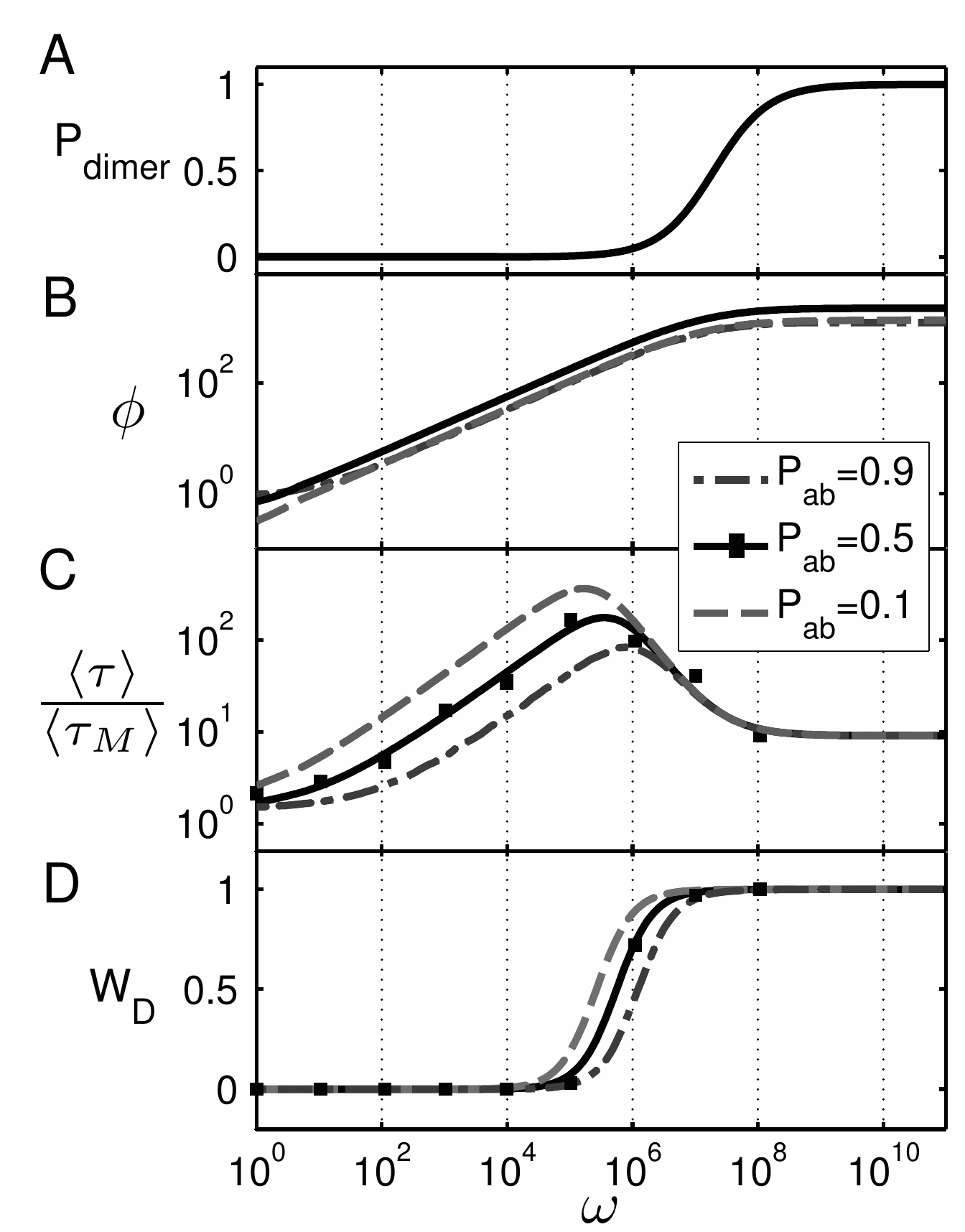}
\caption{
Characterization of the cooperative search process and steady state levels as a function of the cooperativity $\w$ and for different on-state levels $p_{\Ta\Tb}$, given $N=1$ molecule of each TF species. 
(A) Dimerization probability $P_\mathrm{dimer}$ at equilibrium. The dimerization threshold is given by the entropic cost of dimerization and corresponds approximately to the length of the genome $\LG$. (B) The fold-change $\fold$ increases with the cooperativity as $\sqrt{\w}$ below the dimerization threshold and then approaches a maximal value. (C) The cooperative search time $\langle\tau\rangle$ displays a maximum at an intermediate cooperativity. For large $\w$, the search time decreases again and settles at an on-state independent value, corresponding to the dimer search time, cf. Fig.~4. (D) The probability $W_{D}$ that the cooperative target state is reached via the dimer pathway is distinct from $P_{dimer}$ in (A), since independent monomeric search and dimeric search have different time scales. Note that the transition from the monomer to the dimer pathway marks the position of the maximal search time.}
\label{PdimerFoldTsN1}
\end{figure}

Fig.~\ref{PdimerFoldTsN1}B shows the resulting fold-change $\fold=p_{\Ta\Tb}/p_{\Ta}$ for the full model as a function of the cooperativity (on a double-logarithmic scale), with the three curves corresponding to different ON-state levels $p_{\Ta\Tb}$. The fold-change increases monotonously with the cooperativity, roughly as $\fold\sim\sqrt{\w}$, before it saturates at a maximal level that depends slightly on the ON-state level. For $\w\gg 1$, the dependence on the ON-state level $p_{\Ta\Tb}$ is non-monotonous, with a larger $\fold$ for $p_{\Ta\Tb}=0.5$ than for both $p_{\Ta\Tb}=0.1$ and $p_{\Ta\Tb}=0.9$. Much of this behavior can be understood already within the simple approximation of Eqs.~\ref{eq-regulated-recruitment} and \ref{eq-cooperative-activation} as follows: For large $\w$, cooperative binding to the targets becomes dominant in the ON-state, such that the non-cooperative contribution $q_{\Ta}+q_{\Tb}$ in the denominator of Eq.~\ref{eq-cooperative-activation} can be neglected. One then finds that $\fold\approx\sqrt{\w}\sqrt{p_{\Ta\Tb}(1-p_{\Ta\Tb})}$, explaining the behavior in the intermediate $\w$ range of Fig.~\ref{PdimerFoldTsN1}B, i.e. the $\sqrt{\w}$-dependence and the non-monotonous dependence on the ON-state level $p_{\Ta\Tb}$. However, the saturation of the fold-change at very large $\w$ is beyond this simple approximation, which neglects the background DNA and assumes that the TFs hetero-dimerize only on the target. This assumption breaks down in the strongly-interacting regime, as  shown in Fig.~\ref{PdimerFoldTsN1}A, which plots the equilibrium probability to find the TFs as a hetero-dimer. 
Dimers become prevalent in the background when the cooperativity outweighs the entropic cost of dimerization. If the non-specific DNA interaction of monomers is optimized for independent search (see below), the dimerization probability is simply $P_\mathrm{dimer}(\w)=\w/(\w +2\LG)$ (see Section S1). Further increase of $\w$ has no significant effect on the fold-change. Hence, the full model confirms our previous conclusion that a large cooperativity is generally beneficial for the steady-state response, but only up to a value of $\w\sim\LG$. 

Next, we turn to the cooperative search process. We first consider the situation with only one molecule of each type ($\Na=\Nb=1$). Initially, both monomers are unbound. The cooperative search time $\tau$ corresponds to the first point in time when $\Ta$ and $\Tb$ are both occupied. Fig.~\ref{PdimerFoldTsN1}C shows its mean, $\langle\tau\rangle$, as a function of $\w$, for three different ON-state levels $p_{\Ta\Tb}$. Here, the symbols represent simulation results, where the average is taken over a large number of simulation runs (see Section S2 for details), while the solid lines represent an analytical approximation discussed below and in Section S3. Note that $\langle\tau\rangle$ is plotted in units of the monomer search time $\langle\tau_{M}\rangle$, which is defined as the average time needed by a monomer, e.g. of type $\tfa$, to find its target $\Ta$ in the absence of $\tfb$. This kinetic ratio, $\langle\tau\rangle/\langle\tau_{M}\rangle$ is a direct measure of the slow down of cooperative regulation relative to the timescale for independent regulation. 
When the cooperativity is negligible ($\w\approx 1$), Fig.~\ref{PdimerFoldTsN1}C shows that the kinetic ratio is only slightly larger than one. In this regime, the second protein arrives independently and on the same timescale as the first, while each protein is stably bound by itself, such that the first protein typically does not unbind from its target before the second protein arrives. Indeed, the probability of such a ``missed encounter'' depends on the ON-state level $p_{\Ta\Tb}$ and is simply $1-\sqrt{p_{\Ta\Tb}}$ when $\w=1$, which consistently explains the $p_{\Ta\Tb}$-dependence (at fixed $\w=1$) in Fig.~\ref{PdimerFoldTsN1}C. 

With increasing cooperativity $\w$, the cooperative search time also becomes longer. Note that our reference time scale, the monomer search time $\langle\tau_{M}\rangle$, is independent of $\w$, such that the ratio plotted in Fig.~\ref{PdimerFoldTsN1}C shows indeed the $\w$-dependence of the absolute timescale for cooperative search. The slow down scales with the square root of the cooperativity, $\langle\tau\rangle\sim\sqrt{\w}$. This scaling reflects the mechanism underlying the slow down, which is produced by an increasing probability of missed encounters: As the cooperativity is increased, our constraint of a constant $p_{\Ta\Tb}$ implies that a monomer bound to its target becomes less stable and detaches more often before its partner arrives. The cooperative search time is then determined by the number of times a TF must return to its target before finding the other target occupied, which is roughly $1/p_{\Ta}$, the inverse of the probability that a single target is occupied. At intermediate $\w$, this probability scales as $p_{\Ta}\sim\w^{-1/2}$, leading to the observed scaling. 

The increase of the search time $\langle\tau\rangle$ with $\w$ is not indefinite, however, because the relative importance of the dimer pathway increases with $\w$. The contribution of the dimer pathway is shown in Fig.~\ref{PdimerFoldTsN1}D. It displays a sigmoidal form, with a narrow transition region where the cooperative search switches from the monomer mode to the dimer mode. This transition is accompanied by a peak in the cooperative search time in Fig.~\ref{PdimerFoldTsN1}C. Note that this transition occurs at significantly smaller $\w$ values than the transition in the equilibrium probability for hetero-dimerization shown in Fig.~\ref{PdimerFoldTsN1}A. 

\begin{figure}[p]
\centering
\includegraphics[width=8.6cm]{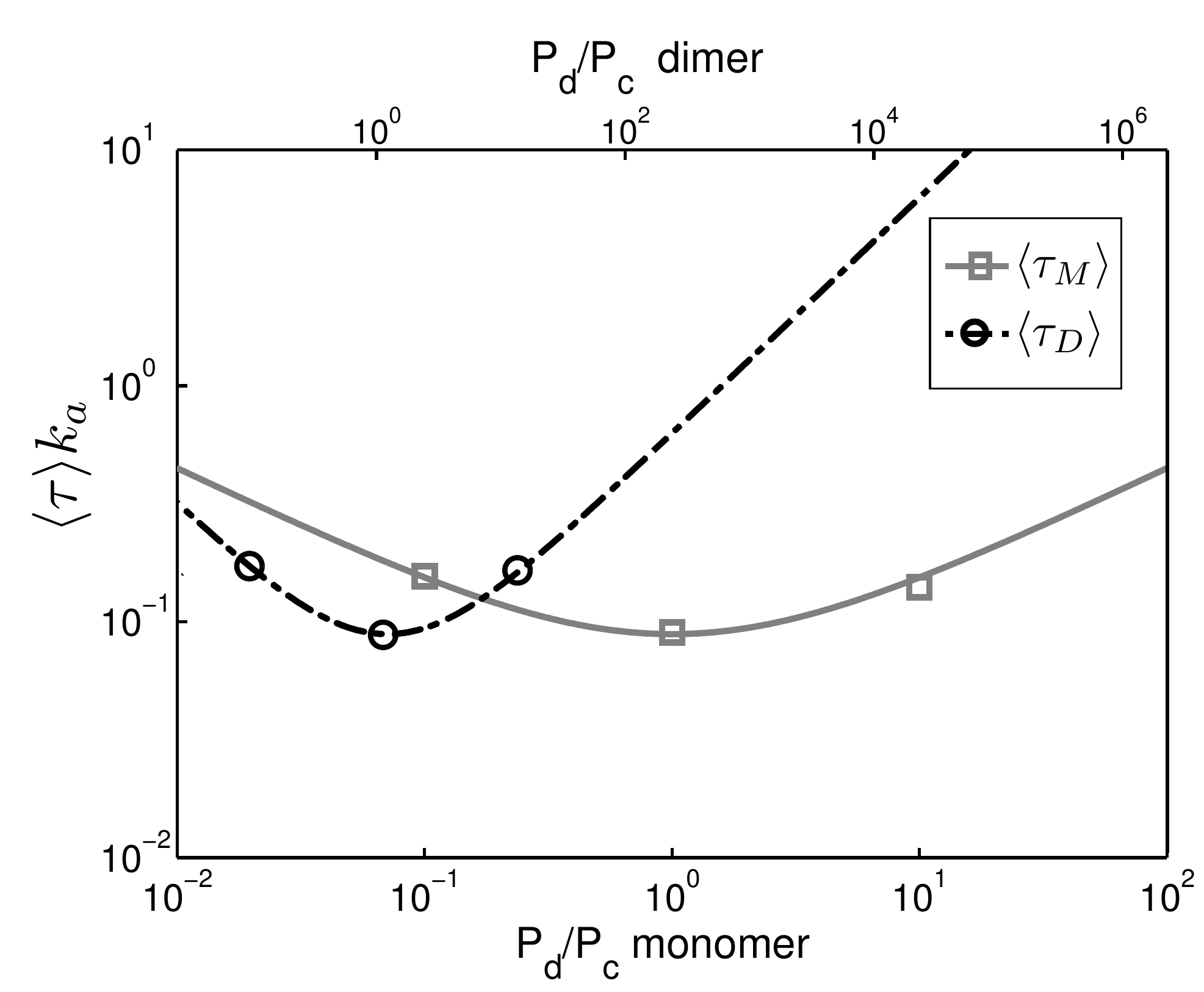}
\caption{Average search times $\langle\tau\rangle$  of a dimer (black) and a monomer (gray) for the target site. The curves are obtained from $\langle\tau\rangle =\sqrt{\pi\LG/16\,\ksl\,\ka}( \sqrt{P_d/P_c} + \sqrt{P_c/P_d})$, which predicts an optimum at a binding ratio of one $P_d/P_c=1$, see Refs.~\cite{Winter_Biochemistry_81b, Slutsky_BiophysJ_04}. For larger binding ratios, TFs spend too much time exploring nearby sites with redundant one-dimensional diffusion, whereas TFs spend too much time unbound in solution when TF-DNA binding is weaker. Since dimers bind DNA more strongly than monomers, the binding ratio $P_d/P_c$ of the dimer (indicated on the top x-axis) is  consistently larger than that of the monomers (bottom x-axis). Hence dimers and  monomers cannot simultaneously operate in the search optimum.}
\label{TsDvsTsM}
\end{figure}

To understand the non-monotonous behavior of the cooperative search time in Fig.~\ref{PdimerFoldTsN1}C, it is instructive to consider the extreme case of a purely dimeric search. Fig.~\ref{TsDvsTsM} shows the purely dimeric search time (black line and circles) as a function of the dimer binding ratio, i.e. the relative probability $P_d/P_c$ to find a dimer on the DNA versus in the cytoplasm (top x-axis). Here, the binding ratio is varied by changing the non-specific binding strength $\Ens$. For comparison, the gray line and squares show the corresponding curve for a monomer (search time for a single target; monomer binding ratio on the bottom x-axis). Both curves display the same qualitative behavior, with the well-known optimum \cite{Winter_Biochemistry_81b, Slutsky_BiophysJ_04} where the respective binding ratio equals one, i.e. the average time spent on the DNA matches the time spent in the cytoplasm. At larger binding ratios, the local 1D search becomes too redundant, whereas at smaller binding ratios TFs spend too large a fraction of their time in solution, not searching. However, the minima of the two search time curves do not coincide, since dimers bind DNA more tightly than monomers. Consequently the protein-DNA interaction cannot be simultaneously optimized for monomer and dimer search. We generally assume that the protein-DNA interaction is optimal for monomers, since single TFs are the basic functional unit for transcription control in bacteria (see below for further discussion). At this point in Fig.~\ref{TsDvsTsM}, the purely dimeric search time is roughly a factor 10 longer than the monomer search time. Returning to Fig.~\ref{PdimerFoldTsN1}, this factor corresponds to the level of the plateau that is reached for very large $\w$ in Fig.~\ref{PdimerFoldTsN1}C. 

We now consider again the intermediate $\w$ range in Fig.~\ref{PdimerFoldTsN1}C. With increasing $\w$ the monomer pathway eventually becomes slower than the dimer pathway, due to the increasing probability of missed encounters. At the same time the dimerized state is increasingly stabilized. Upon dimerization of $\tfa$ and $\tfb$ in the background, it becomes more likely that this dimer localizes the target before it dissociates again into monomers. The increasing predominance of the faster dimer pathway explains the regime where the cooperative search time decreases with $\w$. It also explains why the kinetic monomer-dimer transition in Fig.~\ref{PdimerFoldTsN1}D occurs before the equilibrium monomer-dimer transition in Fig.~\ref{PdimerFoldTsN1}A: even when the dimer fraction has not reached $50 \%$, the dimer pathway can be kinetically dominant. At very large $\w$, the monomer pathway is entirely negligible. The TFs form relatively stable dimers, either already in solution or when bound to non-target sites, subsequently search together for most of the time, and ultimately arrive at the target as a pair. This search mode is independent of the target binding energy and the cooperative search time then becomes independent of $\w$ and equal to the pure dimer search time plotted in Fig.~\ref{TsDvsTsM}. 

The cooperative search kinetics admits an analytical treatment, to quantitatively describe the kinetic competition between the monomer pathway and the dimer pathway. This description takes a coarse-grained view of the problem, with effective transition rates between only four states, as depicted in Fig.~S1. The initial state has both TFs $\tfa$ and $\tfb$ unbound in solution (state 2 in Fig.~S1), from where the proteins either enter the dimer pathway by dimerizing (state 1) at rate $r_2^-$ or one of them independently finds its target site on the DNA (state 3) at rate $r_2^+$. From state 1, the dimer either locates its pair of target sites at rate $r_1^-$ or reverts back to state 2 at the effective dissociation rate $r_1^+$. Along the monomer pathway, from state 3, either the other TF locates its target as well (at rate $r_3^+$), or the waiting TF leaves its target, leading back to state 2 at rate $r_3^-$. In Section S3, we express the six effective rate constants in terms of the parameters of the full model, and then use the mean first passage time formalism to calculate the mean cooperative search time analytically. We have used this approach to obtain the curves in Fig.~\ref{PdimerFoldTsN1}C and D, which agree well with the simulation data.

So far we have focused on the case of a single TF molecule of each type. We now turn to the general case where we have $N_{\tfa}$ molecules of type $\tfa$ and $N_{\tfb}$ molecules of type $\tfb$. If we increase both molecule numbers simultaneously ($N_{\tfa}=N_{\tfb}=N$), mass action drives the monomer-dimer equilibrium towards the dimerized state. Fig.~S2A shows the probability for a molecule to be dimerized, $P_\mathrm{dimer}$, as a function of $\w$, with the different curves corresponding to different $N$ values. As expected, the dimerization threshold of the sigmoidal curves moves to smaller $\w$-values as $N$ is increased. Note that while we have treated the case of exactly one molecule for $N=1$, we keep the number of proteins constant only on average for $N>1$, via the chemical potential in the grand canonical ensemble (see Section S1 for details). This choice is technically motivated, but is also biologically meaningful, since proteins are constantly produced and degraded in cells and their numbers can at best be constant on average. 

Fig.~S2B displays the $N$-dependence of the fold-change $\fold$. In contrast to Fig.~\ref{PdimerFoldTsN1}B, the ON-state level is now kept fixed at $p_{\Ta\Tb}=0.5$ and instead the different curves are for different $N$ values (the fold-change is defined here with respect to the state where $N_{\tfa}=N$ and $N_{\tfb}=0$). For $\w$ below the dimerization threshold, the fold-change is independent of the molecule number $N$. However, as in Fig.~\ref{PdimerFoldTsN1}B, increasing $\w$ does no longer raise the fold-change once the dimerization threshold is reached. As the dimerization threshold decreases with $N$, the fold-change saturates at smaller $\w$ and the maximal $\fold$ decreases as $1/\sqrt{N}$.

The average time required for the parallel cooperative search with $N_{\tfa}=N_{\tfb}=N$ molecules is shown in Fig. \ref{Fig6}C. As in Fig.~\ref{PdimerFoldTsN1}, we have used the monomer search time as the reference time scale, but now scaled by $N^{-1}$, since the expected timescale for the parallel search of $N$ monomers is $\langle \tau_M \rangle/N$. Consequently, the fact that all curves fall on top of each other in the regimes of weak and very strong interaction shows that in these regimes the cooperative search time exhibits the simple $1/N$ scaling, which corresponds to a linear increase of the frequency at which the targets are visited by monomers (in the small $\w$ regime) or by dimers (in the large $\w$ regime). In the intermediate regime, we find a more complex dependence on $N$, indicated by the fact that the curves do not collapse. To understand this dependence, we extend our simplified analytical expression developed above. Under the conditions of interest here, the dimerization equilibrium $P_{dimer}(\w)$ of Fig.~\ref{Fig6}A is reached on a timescale much shorter than the cooperative search. As detailed in Section S3, we can then approximate the search process as a parallel search of $N\cdot P_{dimer}$ dimers and $N\cdot(1-P_{dimer})$ monomers of each kind, resulting in 
\begin{equation}
\langle\tau(\w)\rangle=\frac{1}{N}\cdot \left[ \frac{P_\mathrm{dimer}(\w)}{\langle\tau_{D}\rangle} + \frac{1-P_\mathrm{dimer}(\w)}{\langle\tau_{A,B}(\w)\rangle} \right]^{-1} \;.
\label{Eq_TsofN}
\end{equation}
Here, $1/\langle\tau_{A,B}(\w)\rangle$ is the independent search rate of the monomers, which indirectly depends on $\w$ through the probability of missed encounters, see Section S3, while the dimer search rate is $1/\langle\tau_{D}\rangle$, as in Fig.~\ref{TsDvsTsM}. We used Eq.~\ref{Eq_TsofN} to obtain the lines in Fig.~S2C, which display good agreement with the full simulation, showing that the analytical approximation yields a useful description of the cooperative search kinetics. 

On a more qualitative level, Fig.~S2C shows how the peak in the search time at intermediate $\w$ values is affected by $N$. The peak shifts to smaller $\w$ values with larger $N$, and also becomes less pronounced. From Fig.~S2D, which shows the weight of the dimer pathway in the cooperative search process according to Eq.~S26, we see that the position of the peak remains determined by the switch from the monomeric to the dimeric search mode. The shifted switch to the dimeric search mode, which occurs at smaller $\w$ for larger $N$, also explains the reduction in the peak height: The dimeric search mode takes over before the slowdown of the monomeric search mode becomes dramatic. However, even with hundreds of TF molecules of each species, we still find a peak in the cooperative search time, which divides the $\w$ values into three regimes, as discussed below.

%%%%%%%%%%%%%%%%%%%%%%%%%%%%%%%%%%%%%%%%%%%%%%%%%
\section*{Discussion}
%%%%%%%%%%%%%%%%%%%%%%%%%%%%%%%%%%%%%%%%%%%%%%%%%

We studied the kinetics and the equilibrium statistics of cooperative transcription factor-DNA binding to specific target sites in the genomic background. For our analysis, we considered the dimensionless cooperativity $\w$ as a parameter with a broad range of biochemically feasible values, and sought to identify functional tradeoffs associated with the choice of this value. We focused on the functional context of a signal integration scenario with AND-logic, but the results hold in a similar fashion for a signal transduction scenario, see Fig. \ref{fig1}. From this functional context we derived the central assumption that the average activity of the regulated gene has an optimal level in the ON-state, such that there is a strong selection pressure to maintain this level fixed regardless of the $\w$ value. We satisfied this constraint by compensating changes in $\w$ via the target site binding energy, which is ``programmable'' through the binding site sequence \cite{Gerland_PNAS_02}. Such a compensation has been observed in an analysis of combinatorial promoters, i.e. binding sites tend to deviate from the consensus motif when multiple TFs bind next to each other in the cis-regulatory region \cite{Bilu_GenomeBiol2005}. It is also biologically plausible as it does not interfere with the regulation of genes that are only regulated by one of the TFs or combinatorially with other TFs. 

Given this functional setting, we determined which fold-change in the steady-state activity could be implemented at a given $\w$, and how the kinetic search time depends on $\w$. The fold-change quantifies the discrimination in the promoter output between the states where one or two input signals are present, while the search time is a lower limit to the response time of the regulatory system. The search process has contributions from a monomer and a dimer search pathway, the relative weights of which we determined, again as a function of $\w$. In the regime of weak protein-protein interactions, e.g. $\w<10^{3}-10^{4}$, we found a tradeoff between the kinetics and the steady-state behavior, in the sense that a higher fold-change is associated with a slower response due to a longer assembly time for the protein-complex on the target site. This tradeoff is a consequence of gene activation via the monomer pathway, where individual TFs visit their targets independently and consecutively, possibly dissociating from the target before the cooperative partner arrives (``missed encounters''). In this regime, search time and fold-change both increase as $\sim \w^{1/2}$. At larger $\w$, heterodimers are more stable, increasing the probability that the target is located simultaneously by both proteins (dimer pathway). At the same time, the missed encounters further slow down the independent monomer search, to timescales larger than the dimeric search time. Thus, a transition occurs where the dimer pathway gains weight and the search time decreases again to settle at the purely dimeric search time. 

\subsection*{Assumptions and limitations}

We made a number of simplifying assumptions in our coarse-grained theoretical model. For instance, we assumed that the DNA-binding energy of the dimer is the sum of the binding energies of the monomers. While dimerizing, the monomers may undergo conformational changes that affect the DNA-binding strength \cite{Lefstin_Nature1998}, possibly speeding up the dimeric search. In that case, the peak of the cooperative search time as a function of $\w$ can be even more pronounced than in our model. For simplicity, we assumed identical binding properties of the two TFs $\tfa$ and $\tfb$, however this assumption is without loss of generality and the extension to asymmetric cases is straightforward. We performed the analysis reported here under the assumption of a non-specific background, although we have formulated our model and the theoretical methods to also cover the more general case of a heterogeneous DNA background. A brief analysis of the heterogeneous case has shown that the most significant effect of the heterogeneous background is to slow down the search time in all regimes. For our model, we have also assumed that the cooperativity between the TFs is mediated by a direct interaction. Indirect cooperativity mediated e.g. by DNA bending or looping has the same steady-state properties as direct cooperativity in the low $\w$ regime. However, these indirect mechanisms lead to different steady-state behavior at large $\w$ values and to different kinetics. A detailed analysis of these mechanisms is beyond the scope of this study. 

\subsection*{Biological ramifications and examples}

A central and robust result of our theoretical study is that one can distinguish three qualitatively distinct regimes of TF-TF interaction strengths for transcription regulation: 

(i)~Weak interactions, with a cooperativity $\w<10^{3}-10^{4}$, suffice to implement regulation functions with moderate fold-changes, on the order of 10-fold, in the transcription level. In this regime, the cooperative search time is only moderately elevated above the search time of a single TF (also on the order of 10-fold). In bacteria, where the search time of a single TF molecule is around one minute \cite{Elf_Science_07}, the parallel cooperative search of $10-100$ copies of each TF would then still result in fast responses on the minute timescale. The principal advantage of this regime from a design point of view is that TFs with weak interactions are flexible components, which can be used to control different genes in different ways, alone or cooperatively in various combinations \cite{Buchler_PNAS_03}. Each TF then only needs to be separately optimized for monomeric search (via the non-specific protein-DNA interaction), while cooperative regulation by pairs of TFs is still sufficiently fast. 

(ii)~Interactions of intermediate strength, with $\w$ values in the approximate range of $\w\sim 10^{4}-10^{6}$, lead to cooperative search kinetics that are prohibitively slow, due to an excessive amount of missed encounters. Recent single-molecule experiments have been able to monitor the search process of a single TF {\it in vivo} \cite{Elf_Science_07}. Our prediction of slow cooperative search kinetics could in principle be verified using two-color fluorescence methods. Alternatively, one could measure the transcriptional response time of a synthetically designed, cooperatively regulated gene with a rapid reporter. We also expect that TF-TF interactions within this intermediate regime are avoided by cells. A test of this implication of our study will require a large dataset quantifying a significant subset of the TF-TF interactions in a model organism. To our knowledge, a quantitative high-throughput assay for TF-TF interactions is not yet available and remains as an experimental challenge in the field. Instead, we discuss several specific biological examples below. 

(iii)~Strong interactions, with a cooperativity $\w>10^6-10^7$, allow high fold-changes and a passable response time at the cost of losing combinatorial flexibility: Suppose that each TF signals a different environmental cue, and a set of genes needs to be activated whenever A is present, whereas another, more specialized group of genes is to be activated only if both signals are present. In this situation, a strong heterodimer would not lead to a favorable regulatory design, since the regulation of the unconditional genes by A would be strongly affected by the presence of B. In other words, the strong cooperativity can lead to undesired crosstalk. Nevertheless, this regime of TF-TF interactions is biologically interesting: For instance, strong homodimers can exploit the cooperative stability mechanism to improve the robust function of regulatory circuits \cite{Buchler_PNAS_05}. Also, in cases where the combinatorial flexibility described above is not needed, strong heterodimers could be used to perform a very sharp and AND-like signal integration. This signal integration can be made very rapid by tuning the non-specific protein-DNA interaction of the TFs into a weaker regime, such that the dimer DNA binding ratio $P_d/P_c$ is closer to the optimal value $1$ for search on the DNA. As Fig.~\ref{TsDvsTsM} shows, this would lead to a concomitant decrease of the monomer binding ratio. For TFs that work in this regime, we therefore expect that monomers spend less than $50\,\%$ of their time bound on DNA. So far, the DNA binding ratios of transcription factors have not been assayed on a large scale. Such an experiment would yield interesting clues about the design and the mode of operation of these TFs. 

Finally, we discuss biological examples. Currently, 383 operons in {\it E. coli} are known to be transcriptionally regulated by two or more TFs (see Section S4). However, it is not known what fraction of these regulatory interactions involves cooperative protein-DNA binding. One well-studied case of co-dependent activation is the \textit{melAB} promoter, where CRP and MelR bind cooperatively and activate transcription \cite{Wade_EMBO_2001}. The interaction of CRP and MelR occurs via a weak surface contact and the binding of either is found to be reduced if the binding of the partner is impeded. In the presence of both, the transcription rate is tenfold increased \cite{Wade_EMBO_2001}. This case is a good example for our regime (i). It is interesting to note that the binding sites of CRP and MelR in the \textit{melAB} promoter display a relatively poor match to the consensus sequence, which is consistent with our assumption that the target binding energies are evolutionarily tuned. Also, CRP is a well known global regulator that controls many other genes in different ways, and hence the combinatorial flexibility achieved with a small cooperativity $\w$ appears to be amply exploited by {\it E. coli}. Other examples of prokaryotic co-activation are the ansB promoter, activated by CRP and FNR \cite{Scott_MolMicroBio_1995}, and the activation of the \textit{mapEP} promoter by CRP and MalT \cite{Richet_EMBO_2000, Richet_EMBO_1996}. More generally, the regime (i) corresponds to the regulated recruitment mechanism for transcription regulation \cite{Ptashne_GenesSignals2001}, which appears to be widely used in eukaryotes. Indeed, the case of the \textit{melAB} promoter described above has been described as a bacterial version of eukaryotic enhanceosomes \cite{Wade_EMBO_2001}. 
A prokaryotic example for regime (iii) may be the RcsA/RcsB heterodimer which is required to activate capsule expression through the RcsF phosphorylation cascade \cite{Majdalani_JBact_2005}. Interestingly, RcsB can also from homodimers and regulate the transcription of other genes by itself, suggesting that this TF may be optimized to always search and function as a dimer (homo- or heteromeric).

%%%%%%%%%%%%%%%%%%%%%%%%%%%%%%%%%%%%%%%%%%%%%%%%%
\section*{Conclusion}
%%%%%%%%%%%%%%%%%%%%%%%%%%%%%%%%%%%%%%%%%%%%%%%%%

We reported a biophysical analysis of the design principles for TF-TF interactions. The exploration of our theoretical model lead us to two functionally favorable regimes for the cooperativity $\w$, corresponding to weak, glue-like promiscuous interactions and very strong heterodimerization, respectively. Cells appear to implement both favorable regimes, but in different biological contexts. On the other hand, our model predicts that the search kinetics will be prohibitively slow at intermediate $\w$ values, at least when the protein copy number is small as is typically the case for transcription factors. Hence the intermediate $\w$-regime appears undesirable in this functional context. This prediction could be tested with experimental approaches from single-molecule biophysics. Currently, there is only limited biochemical data available for the cooperativity values involved in transcription regulation, typically from {\it in vitro} experiments with selected DNA-binding proteins. Once more data becomes available, it will be interesting to see whether the intermediate $\w$-regime is indeed avoided.

\section*{Acknowledgements} 
We thank Nicolas Buchler and Karin Schnetz for helpful comments. We especially thank Terry Hwa for stimulating discussions in the initial phase of this study. 
Financial support from the DFG and the German Excellence Initiative via the ÔNano-Initiative Munich (NIM)Õ is gratefully acknowledged. This work was partially supported by the Spanish Ministry of Education, grant number FPU-AP-2007-00975.

\appendix
\newpage

\setcounter{figure}{0}
\setcounter{equation}{0}
\setcounter{section}{0}
\renewcommand\thefigure{S\arabic{figure}}
\renewcommand\thetable{S\arabic{table}}
\renewcommand\theequation{S\arabic{equation}}    
\renewcommand\thesection{S\arabic{section}}    

\noindent 
{\huge \bf Supporting Material\\}

\section{Exact calculation of steady-state activities}

\paragraph{Single TF molecules.}
We first treat the case where the cell contains only a single molecule of each TF species, $\Na=\Nb=1$. The equilibrium statistics of the system is described by the canonical ensemble of statistical physics. The appropriate Boltzmann weight for a single TF binding to one of $\LG$ sites in a non-specific DNA background is $q_{ns}=\exp(-\Ens)$ (see below for the most general case with an arbitrary background and larger TF numbers). For a purely non-specific background and $S=\Vc/\Vtf\gg\LG$ unbound states, the partition function is 
\begin{eqnarray}
 \Zback & = &\LG(\LG-2\Ltf)\qns^2 +S^2 \nonumber \\
  & + & 2 S \LG \qns \nonumber \\
  & + & \w \left(\LG\qns^2  + S\right) \;.
\label{Zbackcanonic}
\end{eqnarray}
The first three terms describe the non-interacting states, where $\tfa$ and $\tfb$ are either separately bound to the DNA to non-adjacent sites, or both are free but not dimerized, or one is DNA-bound and the other is free. The fourth term corresponds to the states where $\tfa$ and $\tfb$ are dimerized, either on the DNA or unbound. The fraction of dimers in the background corresponds to the ratio of the weights of the dimerized states to the weight of all possible states, $\w \left(\LG \qns^2 +S\right)/\Zback$. Rewriting this expression in terms of the monomer DNA binding ratio $\alpha=P_d/P_c= \qns \LG/S$, one obtains 
\begin{equation}
P_\mathrm{dimer}(\alpha,\w)=\frac{\w}{\w + \left( S(\alpha^2+1) +2\alpha\right)/(\alpha \qns +1)} \;.
\end{equation}
For a binding ratio of one, i.e. when the monomers are optimized for independent search, $P_\mathrm{dimer}(\w)=\w/(\w +2\LG)$, which is the case plotted in Fig.~3A. Here, a dimerization probability of $0.5$ is reached at $\w_{1/2}=2\LG$, while we would have $\w_{1/2}=S$ for $\alpha\rightarrow 0$ and $\w_{1/2}=\LG$ for $\alpha\rightarrow \infty$.

Eq.~\ref{Zbackcanonic} provides the binding-statistics on non-target states. To study the full system, we add the target states with weights $q_T=\exp(-\Et)$ for the full partition function 
\begin{equation}
\Zt=\Zback+[2(\LG-\Ltf-1)\qns + S]q_T +\w q_T^2 \;,
\label{Ztcanonic}
\end{equation}
where the second term is the weight of a single occupied target and the third term is the weight for both targets to be occupied simultaneously. Hence the double target occupation probability is $p_{\Ta\Tb}=\w q_T^2/\Zt$. This equation can be interpreted as a quadratic equation for $q_T$ for given values of $p_{\Ta\Tb}$ and $\w$ (since our analysis assumes a fixed $p_{\Ta\Tb}$ corresponding to the optimal occupation-probability of the targets in the ON-state). Hence we obtain an explicit expression for $\Et(\w,p_{\Ta\Tb})$ (not shown), which we use throughout this paper to determine $\Et$ for the kinetic model and stochastic simulations in the $\N=1$ case. Furthermore, to calculate the fold-change $\fold=p_{\Ta\Tb}/p_{\Ta}$ at a given $\Et(\w,p_{\Ta\Tb})$ we determine the probability of single TF target binding $p_{\Ta}$ in the absence of a partner. By calculating the partition function for a system of a single TF, we find
\begin{equation}
p_{\Ta}=p_{\Tb}=\frac{\e^{-\Et(\w,\,p_{\Ta\Tb})}}{(\LG-1)\qns+S+\e^{-\Et(\w,\,p_{\Ta\Tb})}} \;.
\end{equation}
For small $\w$, this probability scales as $\sim\w^{-1/2}$.

\paragraph{Multiple TF molecules.}
For the case of multiple TF molecules, we calculate the exact equilibrium statistics of our full model using the standard transfer matrix approach from statistical physics, see e.g. \cite{Schwabl_2006, Teif_NucleicAcidsRes2007}. The calculation is based on the grand canonical ensemble, i.e. the average copy numbers $\Na$, $\Nb$ of the proteins $\tfa$ and $\tfb$ are set by the corresponding chemical potentials $\mua$, $\mub$. The total partition function $Z$ of the complete system then factorizes, 
\begin{equation}\label{Z}
Z=\Zd\Zc \;,
\end{equation} 
into a product of a ``DNA partition function'' $\Zd$ involving only the DNA-bound states of the TFs and a ``cytosol partition function'' $\Zc$ involving only the unbound states (the factorization is possible because DNA-bound TFs do not interact with unbound TFs and because the TF numbers are not conserved in the grand canonical ensemble). Due to the low TF concentrations in the cytosol, steric exclusion between unbound TFs is negligible, and $\Zc$ takes the simple form 
\begin{eqnarray}\label{Zc}
\Zc & = & \left( 1 + \e^{\mua} + \e^{\mub} + \w \,\e^{\mua+\mub} \right)^S \;,
\end{eqnarray}
where $S\gg\LG$ is the number of solvent states (i.e. the ratio of the cell volume to a characteristic TF volume, $S=\Vc/\Vtf$) and the statistical weight for an unoccupied solvent state is one. For the calculation of the DNA partition function $\Zd$, we do take the steric exclusion of DNA-bound TFs into account. The number of base pairs covered by a single TF molecule is denoted by $\Ltf$. Each base pair $i=1\ldots\LG$ on the genome can then be in one of $2\Ltf +1$ states: In state 0, the base pair is not covered by a TF. In state 1, it is the leftmost contact position of a TF of type $\tfa$, in state 2 it is the second leftmost contact position, and so on, up to state $\Ltf$ corresponding to the rightmost contact position of $\tfa$. States $\Ltf+1$ up to $2\Ltf$ are analogous for $\tfb$. The transfer matrix $Q_{i}$ describes the statistical coupling between the states of the neighboring DNA positions $i$ and $i+1$. Each $Q_{i}$ is a square matrix of dimension $2\Ltf+1$, defined such that the partition function is equal to the trace of the (ordered) product of all transfer matrices, 
\begin{equation}\label{Zd}
Z_d=Tr\left(\prod_{i=1}^{\LG} Q_{i}\right) \;,
\end{equation}
for a circular DNA with $\LG$ basepairs (for a linear DNA molecule, the trace operation would have to be replaced by multiplication of a row vector from the left and a column vector from the right, with the vector components properly chosen to enforce the boundary conditions). Let us denote by $[Q_{i}]_{ss'}$ the element in row $s$ and column $s'$ of the transfer matrix at position $i$. It takes on a non-negative value, which corresponds to the conditional statistical weight of finding position $i$ in state $s'$, provided that position $i-1$ is in state $s$. Thus, each $[Q_{i}]_{ss'}$ is a Boltzmann factor that accounts for the contribution to the total configurational energy that stems from position $i$ and its interaction with position $i+1$. The Boltzmann factor is zero, if the two states are incompatible (overlapping TFs or a single TF binding to non-contiguous basepairs). The non-zero entries of $Q_{i}$ contain the protein-DNA binding energy landscapes $\Eia{i}$ and $\Eib{i}$, the cooperativity $\w$, and the chemical potentials. For illustration, we show the transfer matrix $Q_{i}$ for TFs of length $\Ltf=2$, 
\begin{equation}\label{TransferMatrix}
Q_{i}=\left(
\begin{array}{ccccc}
 1 & \e^{-\Eia{i}+\mua}  & 0 & \e^{-\Eib{i}+\mub} & 0  \\
 0 & 0 & 1 & 0 & 0  \\
 1 & \e^{-\Eia{i}+\mua} & 0 & \w\,\e^{-\Eib{i}+\mub} & 0  \\
 0 & 0     & 0  & 0 & 1  \\
 1 & \e^{-\Eia{i}+\mua}  & 0 & \e^{-\Eib{i}+\mub} & 0  \\
\end{array} \right) \;.
\end{equation}
The entries with value one reflect the mere compatibility of neighboring states without an energetic contribution (e.g., when position $i-1$ is in state $1$, position $i$ must be in state $2$, and there is no additional energy contribution to take into account). Note that we assume a directional interaction between the TFs $\tfa$ and $\tfb$  (the attractive contact only occurs when $\tfb$ is bound directly downstream from $\tfa$). 

From the partition function (\ref{Z}), we can obtain exact expressions for the occupation probabilities of DNA sites by differentiation. For instance, the probability that a TF molecule of type $\tfa$ is bound to the site starting at position $i$ on the DNA is 
\begin{equation}
p^{\tfa}_i=-\frac{\partial}{\partial \Eia{i}} \log Z  \;.
\label{monomer-occupancy}
\end{equation}
The derivative is straightforward to evaluate explicitly, leading to an expression of the form $p^{\tfa}_i=Z_{d}'/Z_{d}$, where the restricted partition function $Z_{d}'$ has the same form as (\ref{Zd}), but with a projection matrix next to $Q_{i}$ inside the trace. This exact expression is easily computed numerically, in particular when large parts of the binding energy landscapes $\Eia{i}$ and $\Eib{i}$ are flat (equal to the non-specific binding energy $\Ens$), since large parts of the product in (\ref{Zd}) then reduce to matrix powers (which are quickly calculated via diagonalization). 
Similarly, the probability of cooperative binding at site $i$ is calculated starting from the expression 
\begin{equation}
p^{AB}_i=\frac{\partial^{2}}{\partial \Eia{i}\partial \Eib{i+\Ltf}} \log Z  \;, 
\label{dimer-occupancy}
\end{equation}
where the derivatives enforce that a $\tfb$ molecule is bound directly adjacent to the $\tfa$ molecule, such that together they cover the DNA positions from $i$ to $i+2\Ltf-1$. Finally, the average number of TF molecules in the system at given values of the chemical potentials $\mua$, $\mub$ are obtained by summing over the occupation numbers of all states, e.g. 
\begin{equation}
\Na=\sum_{i=1}^{\LG} p_i^A + \frac{S (\e^{\mua}+\w \,\e^{\mua+\mub})}{Z_c} \;.
\end{equation}
Similarly, the average number of dimers in the system is 
\begin{equation}
N_\mathrm{dimer}=\sum_{i=1}^{\LG} p_i^{AB} + \frac{S\,\w \,\e^{\mua+\mub}}{Z_c} \;,
\end{equation}
from which the fraction of dimers, $P_\mathrm{dimer}(\w)=N_\mathrm{dimer}/N$, in Fig.~6A is computed. The fold-change $\fold$ in Fig.~6B is calculated as the ratio of the dimer occupancy (\ref{dimer-occupancy}) at the target site pair in the presence of both TFs ($\mua=\mub\equiv\mu$ such that $\Na=\Nb\equiv N$) to the monomer occupancy (\ref{monomer-occupancy}) at its target site when only one TF is present ($\mua$ chosen such that $\Na=N$ while $\mub$ is set to a large negative value such that $\Nb\approx 0$). 

The above framework can be used to calculate any equilibrium observable exactly for our full model and it also provides a reference point for our kinetic simulations, which produce equilibrium values in the long-time average. However, it is also useful to derive a simple approximation to the exact solution of the multiple TF molecule case, which still incorporates the effect of a (nonspecific) DNA background, but neglects steric exclusion between the TFs in the background. Assuming $\e^{\Ens}\ll 1$ and $\N\ll\LG$, and again taking a DNA binding ratio of one, such that $\S\e^{\Ens}=\LG$, we find 
\begin{equation}
\qns \equiv \e^{-\Ens+\mu}\approx \frac{\sqrt{1+\frac{\N}{\LG}\w}-1}{\w} \;,
\label{Eq_qnsapprox}
\end{equation}
which leads to the background dimerization fraction
\begin{equation}
P_\mathrm{dimer}(\w) \approx 1- \frac{2\LG}{\N\w}\left(\sqrt{1+\frac{\N\w}{\LG}} - 1 \right) 
\label{Eq_Pdimerapprox}
\end{equation}
that we use in Eq.~4 of the main text for the approximative form of the cooperative search time.

\section{Stochastic simulation of cooperative search kinetics}

To study the cooperative search process within the full reaction scheme of Fig.~2B, we implemented a kinetic Monte Carlo simulation based on the standard Gillespie algorithm. For our simulations, we used fixed numbers, $\Na$ and $\Nb$, of $\tfa$ and $\tfb$ molecules (i.e., any equilibrium values computed in these simulations correspond to thermodynamic averages in the canonical ensemble). The state of the system is specified by the state of each TF molecule, which can be either free or dimerized in solution, or bound to the DNA at position $p$. The simulations generate stochastic continuous-time trajectories in this discrete state space. Each simulation step consists of one of the moves depicted in Fig.~2B, however the set of available moves depends on the current state of the system. In particular, moves that would violate the steric constraint that each DNA basepair can be be in contact with only a single TF molecule cannot be chosen. Thus, TF molecules can, for instance, not change the order at which they are bound along the DNA solely via sliding moves.

To measure the average cooperative search time $\langle\tau\rangle$, we perform 100 simulations for each set of model parameters. Each simulation run is initialized in the state where all molecules are unbound (this mimics the condition of a cell prior to receiving a signal that triggers allosteric activation of TF-DNA binding), and terminated once the the two adjacent target sites are both occupied simultaneously. The data points in Fig.~3C, Fig.~4, and Fig.~6C correspond to the simulation time averaged over the 100 runs. Another observable of interest here is the relative contribution of the dimer pathway to the search process, as shown in Fig.~3D and Fig.~6D. This observable corresponds to the fraction of simulation runs where the final state is reached by a dimer move, such that both targets simultaneously become occupied by their cognate TF molecule.

\section{Analytical description of the cooperative search kinetics}

Here, we develop a simplified analytical description of the cooperative search kinetics, which distinguishes only the target occupation states and the two search modes (dimeric vs. monomeric). As shown in Fig.~S1, this description corresponds to a kinetic scheme with four states and six effective rates. The scheme amounts to two competing Michaelis-Menten type processes which lead to the same final state. The initial state 2 corresponds to the state of our TF-DNA system where both proteins are unbound. From there, the target state can either be reached via state 1 (dimer pathway) or via state 3 (monomer pathway). The dimer pathway is kinetically characterized by the effective dimerization rate $r_2^-$, the effective dissociation rate $r_1^+$, and the dimer search rate $r_1^- \equiv 1/\langle\tau_D \rangle$. Similarly, the monomer pathway is characterized by the three rates $r_2^+$, $r_3^-$, and $r_3^+$. Since state 3 does not distinguish whether $\tfa$ or $\tfb$ is bound, the rate $r_2^+\equiv 2/\langle\tau_M \rangle$ is twice the monomer search rate. In contrast, the rate $r_3^+\equiv 1/2\langle\tau_M \rangle$ corresponds to only half the search rate of a monomer because one target is already occupied and the other target is accessible from one side only. Finally, $r_{3}^{-}$ is the total rate at which a monomer dissociates from its target, either via sliding or unbinding. 

We can express the three remaining undetermined rate constants $r_2^-$, $r_1^+$, and $r_{3}^{-}$
in terms of our underlying model parameters. For arbitrary binding energy landscapes, the effective dimerization rate is 
\begin{equation}
r_2^- = \sum_{i\neq \Ta, \Tb} \left[ \left( k^{\tfa+}_{i}+k^{\tfb-}_{i+\Ltf+1} \right) p_i^\tfa \, p_{i+\Ltf+1}^\tfb 
+ \ka \, p_i^\tfa P_c^\tfb + \ka \, p_i^\tfb P_c^\tfa  \right]  + \ka P_c^\tfa P_c^\tfb\;,
\end{equation}
where we have used the equilibrium probabilities introduced above in section A of `Methods', and $P_c^\tfa$, $P_c^\tfb$ denote the equilibrium probabilities for the TFs to be unbound in solution. The rates $k^{\tfa+}_{i}$ and $k^{\tfa-}_{i}$ denote the forward and backward sliding rates from position $i$, see section `Full model'. Using our approximations from section A for a non-specific background, we find the simpler form for the effective dimerization rate
\begin{equation}
r_2^-= \left( \frac{2\,k_\mathrm{sl}}{\LG} -\ka \right)P_d^2  + \ka \;,
\label{Eq_DimDissRateApprox}
\end{equation}
where $P_{d}=1- P_c^\tfa = 1 - P_c^\tfb$ is the probability to find a TF molecule bound to DNA. Similarly, the effective dissociation rate has the general form 
\begin{equation}
r_1^+ = \sum_{i\neq \Ta} \frac{p^{AB}_i}{\w}\left(k^{\tfa, \mathrm{off}}_{i} + k^{\tfb, \mathrm{off}}_{i+L} + k^{\tfa-}_{i}+k^{\tfb+}_{i+L} \right) + k_\mathrm{d} \, P_c^{AB} \;,
\end{equation}
where $k^{\tfa, \mathrm{off}}_{i}$ denotes the site-specific DNA-unbinding rate for $\tfa$ and $P_c^{AB}$ is the probability to find the two TFs dimerized in solution. The simplified effective dissociation rate for a non-specific background is  
\begin{equation}
r_1^+ =\frac{2 P_d^{AB} }{\w} \left( k_\mathrm{sl} + k_\mathrm{off}\right) + k_\mathrm{d} \, P_c^{AB}\;,
\end{equation}
where $P_d^{AB}$ is the total probability to find the TFs non-specifically bound to the DNA as a heterodimer. 
Finally, the total rate for monomer loss from a target is  
\begin{equation}
r_3^- = k_{\mathrm{off},\Ta} + 2\,k_{\mathrm{sl},\Ta} \;.
\end{equation}
where the index $\Ta$ indicates that these are unbinding and sliding rates from the target site, which are slower than their bulk counterparts by the additional Boltzmann factor corresponding to the energy difference between the non-specific binding energy and the target binding energy, see section `Full model'. 

With these rates, the average assembly time of the two TFs on the double target corresponds to the mean first passage time (MFPT) of a random walker hopping between the four sites at the given site-dependent jump rates. The random walker starts at site 2 and terminates on the target site. We use the standard MFPT formalism as described, for instance, in Ref.~\cite{Gardiner_2004} to calculate this cooperative search time. The general formula for the MFPT $\langle \tau(M) \rangle$ starting from site $M$ on a linear lattice with $N+1$ sites, with the two boundary sites $0$ and $N$ both absorbing, is
\begin{equation}
\langle \tau(M) \rangle= W(M) \sum_{m=1}^{N-1} \sum_{n=1}^{m} \frac{1}{r_n^+} \prod_{j=n+1}^m \frac{r_j^-}{r_j^+}  - \sum_{m=1}^{M-1}\sum_{n=1}^{m}\frac{1}{r_n^+}\prod_{j=n+1}^m \frac{r_j^-}{r_j^+} \;, 
\end{equation}
where $W(M)$ is the total probability to exit to site $N$, 
\begin{equation}
W(M) = \frac{1+ \sum\limits_{m=1}^{M-1} \prod\limits_{j=1}^{m} \frac{r_j^-}{r_j^+}}{1+ \sum\limits_{m=1}^{N-1}\prod\limits_{j=1}^m \frac{r_j^-}{r_j^+}} \;.  
\end{equation}
For the problem at hand, we have $N=4$ and $M=2$. 
Defining the Michaelis-Menten-type constant $K_{1}=(r_1^{-}+r_1^{+})/r_2^{-}$ for state 1 and $K_{3}=(r_3^{+}+r_3^{-})/r_2^{+}$ for state 3, we can rewrite the cooperative search rate, i.e. the inverse average search time, in the compact form 
\begin{equation}
\frac{1}{\langle \tau \rangle} = \frac{K_{1}\,r_3^{+}+K_{3}\,r_1^{-}}{K_{1}+K_{1}K_{3}+K_{3}} \;,
\label{Eq_ExactCooperativeSearchTime}
\end{equation}
which is the expression used to obtain the lines in Fig.~3C. In the limit where $r_2^-$ vanishes, this reduces to the average search rate for two independent monomers,  
\begin{equation}
\frac{1}{\langle \tau_{A,B} \rangle} = \frac{r_3^+}{1+ K_{3}} \;.
\label{Eq_ExactIndependentSearchTime}
\end{equation}
Using the relation  $2\, r_{3}^{-} p_{\Ta}=r_{2}^{+} (1-p_{\Ta})$, we can rewrite the corresponding search time in the form 
\begin{equation}
\langle \tau_{A,B} \rangle = \left(\frac{5}{2} + \frac{1-p_{\Ta}}{p_{\Ta}} \right) \, \langle\tau_M\rangle  \;,
\end{equation}
which best explains the effect of missed encounters where $1/p_{\Ta}$ is the average number of times a TF must return to the target before finding the other target occupied. In the small $\w$ regime the cooperative search process corresponds to an independent monomer search and $\langle\tau\rangle \approx \langle \tau_{A,B} \rangle$. Given that $p_{\Ta}\sim \w^{-1/2}$, this form also explains the $\langle\tau\rangle \sim \sqrt{\w}$ scaling of the search time at small cooperativities.

We can further simplify Eq.~\ref{Eq_ExactCooperativeSearchTime} by noting that the average search time is virtually identical (in the parameter regime considered here) when the search begins in state 1 instead of state 2. With state 1 as the initial state, we find  
\begin{equation}
\langle \tau \rangle =\left( r_1^- P_\mathrm{dimer} + \frac{1}{\langle \tau_{A,B} \rangle }(1-P_\mathrm{dimer})\right)^{-1} \;.
\label{Eq_mfpt_InTermsOfCurrents}
\end{equation}
The first term corresponds to the dimer pathway, while the second term corresponds to the monomer pathway. As expected, the contribution of either pathway depends on the dimerization probability and on the search rate of the respective mode. It follows that the relative weight of the dimer pathway can be written as
\begin{equation}
W_{D}(\w)=\frac{P_\mathrm{dimer}(\w)\,r_1^-}{P_\mathrm{dimer}(\w)\,r_1^- + (1-P_\mathrm{dimer}(\w)) \langle \tau_{A,B} \rangle^{-1}} \;,
\label{eq:WD}
\end{equation}
which was used to obtain the lines in Fig.~3D.
It is straightforward to generalize these equations also to the case of $N>1$, where the dimerization probability $P_\mathrm{dimer}(\w,N)$ becomes a function of both $\w$ and $N$, and the search rate for each mode increases by a factor of $N$: $ r_1^- \rightarrow N r_1^-$ and $\langle \tau_{A,B} \rangle \rightarrow \langle \tau_{A,B} \rangle /N$. In this case we obtain Eq.~4 from the main text which is used to obtain the analytical curves in Fig.~S2C. Using the dimerization probability $P_\mathrm{dimer}(\w,N)$, we also extend Eq.~\ref{eq:WD} to the case of $N>1$, to obtain the curves in Fig.~S2D.

\section{Additional notes}

To obtain an estimate of the number of {\it E. coli} operons which are regulated by two or more transcription factors, we perused the ``RegulonDB'' database \cite{RegulonDB}. At the time of writing, this database lists 370 {\it E. coli} operons as regulated by a single transcription factor, while 383 operons are listed as regulated by two or more transcription factors (188 of these are believed to be regulated by exactly two transcription factors).

\begin{figure}[p]
\centering
\includegraphics[width=9cm]{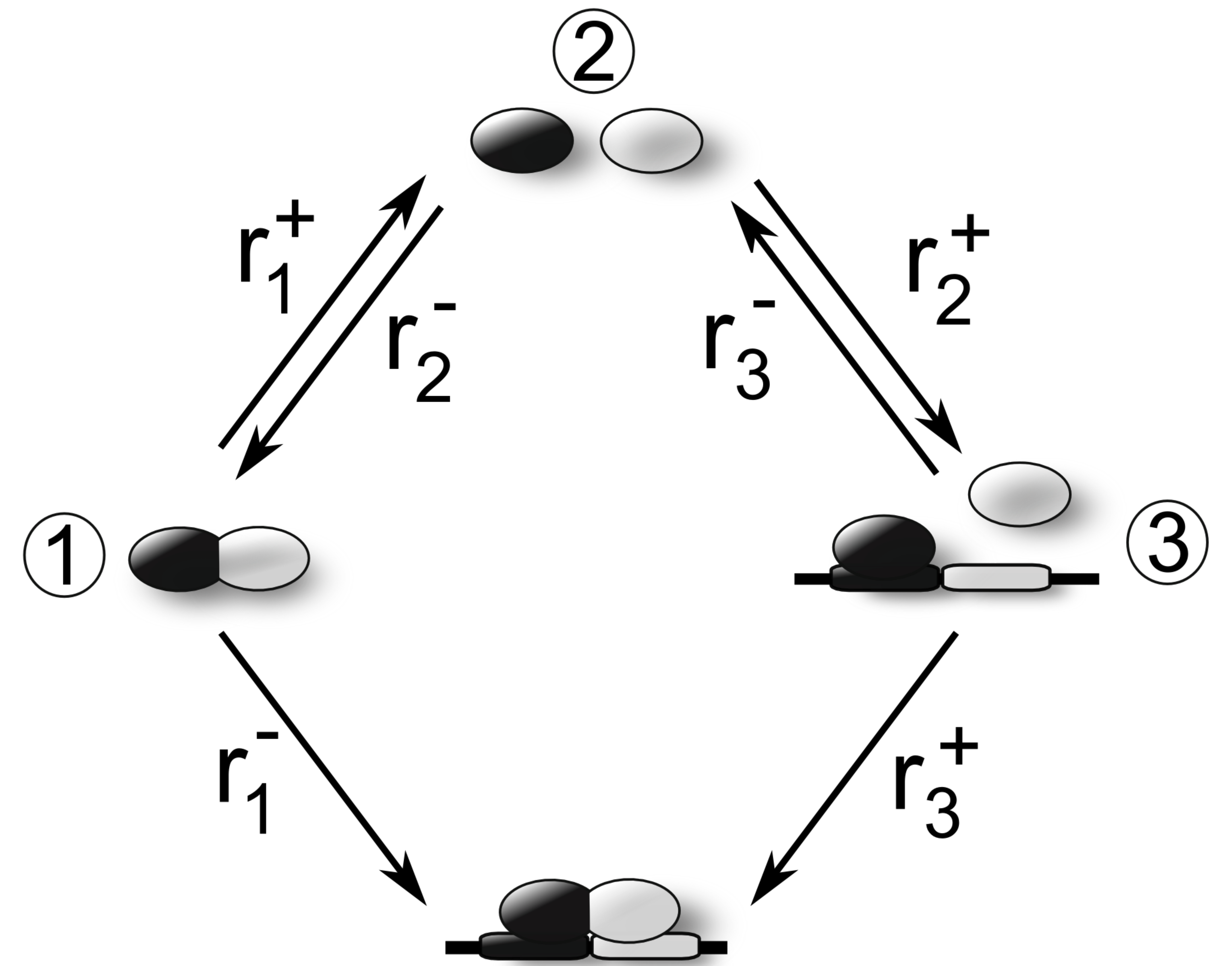}
\caption{Simplified model used to  calculate the mean cooperative search time analytically. In this model only the different target occupation states and the dimeric vs. monomeric search modi are distinguished. The rates $r_1^-$ and $r_3^+$ correspond to the search rates of dimers or monomers respectively, whereas $r_2^-$ and $r_1^+$ are the total rates at which a dimerization or a dissociation occur in the dimeric or monomeric state, respectively. The rate $r_3^-$ refers to the total rate at which a monomer leaves its target, either by sliding away or by dissociating from it.
}
\label{Fig_4state}
\end{figure}

\begin{figure}[p]
\centering
\includegraphics[width=9cm]{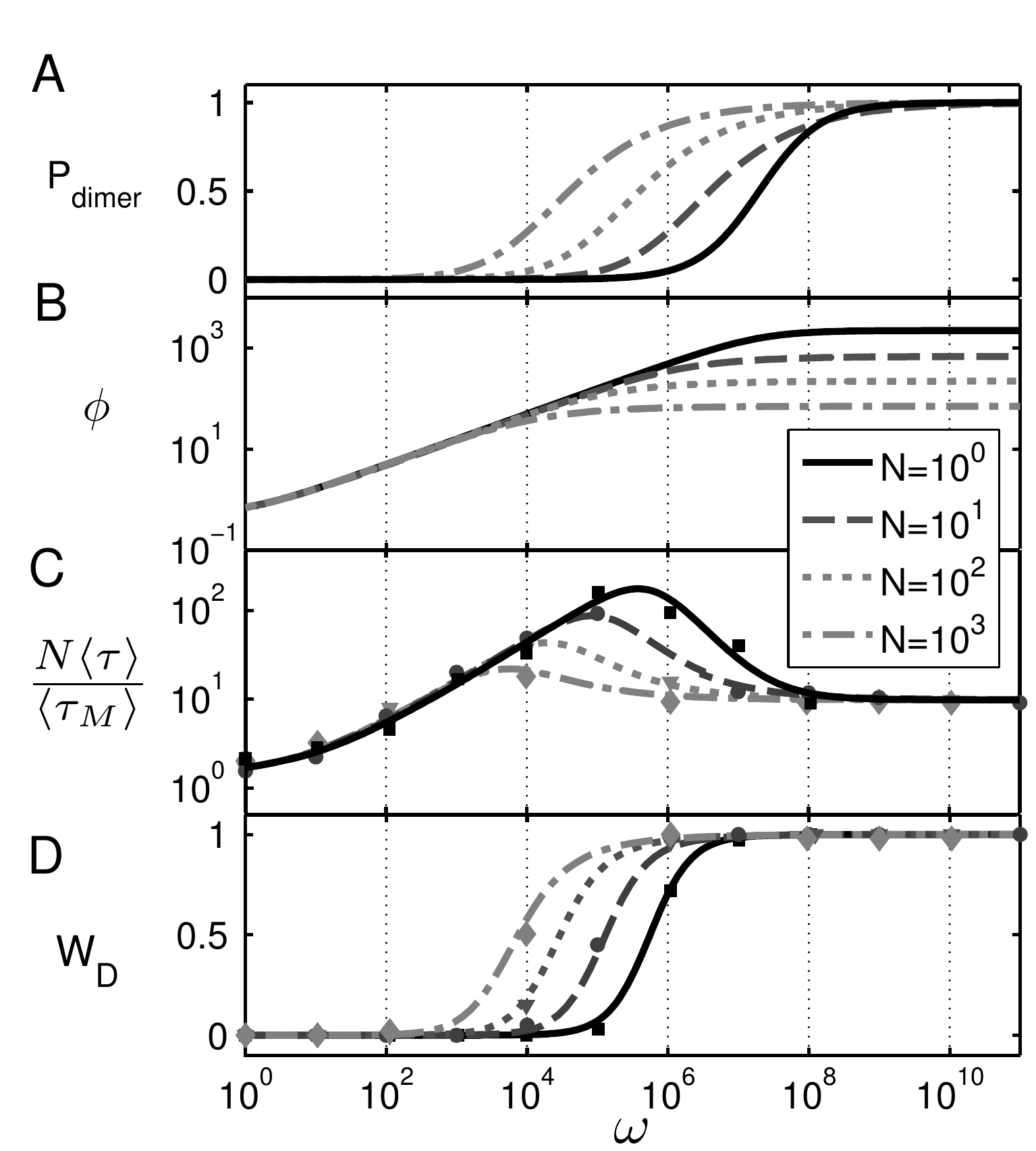}
\caption{Cooperative search times and steady state levels as a function of $\w$, given different TF copy numers $N=1,10,100$, and $1000$ at a fixed $p_{\Ta\Tb}=0.5$. (A) The dimerization threshold decreases with increasing TF concentrations whereas the foldchange (B) is independent of the TF number in the monomeric regime. The maximal foldchange is reached at the dimerization threshold, which decreases with the TF concentration, such that the maximal foldchange in (B) decreases as well. The search time (C) scales as $1/N$ in the purely monomeric and purely dimeric regime. In the intermediate regime, the maximal search time decreases stronger than $1/N$, as the onset of the dimeric pathway (shown in D) moves to lower cooperativities.
}
\label{Fig6}
\end{figure}

\end{document}